\documentclass[preprint]{aastex}
\usepackage{bm}
\usepackage{graphicx}
\usepackage{times}
\usepackage{amssymb,amsmath}

\def\araa{ARAA}

\def\mnras{MNRAS}
\def\jcap{JCAP}

\def\apj{ApJ}

\def\prd{PRD}

\def\beq{\begin{equation}}
\def\eeq{\end{equation}}
\def\bey{\begin{eqnarray}}
\def\eey{\end{eqnarray}}
\def\bfig{\begin{figure}}
\def\efig{\end{figure}}

\def\lsim{\mathrel{\raise.3ex\hbox{$<$\kern-.75em\lower1ex\hbox{$\sim$}}}}
\def\gsim{\mathrel{\raise.3ex\hbox{$>$\kern-.75em\lower1ex\hbox{$\sim$}}}}

\begin{document}
\title{Cluster--Void Degeneracy Breaking: Dark Energy, {\it Planck}, and the Largest Cluster \& Void}

\author{Martin Sahl\'en, \'I\~nigo Zubeld\'ia, and Joseph Silk\altaffilmark{1,2}}
\affil{BIPAC, Department of Physics, University of Oxford}
\affil{Denys Wilkinson Building, 1 Keble Road, Oxford OX1 3RH, UK}
\email{msahlen@msahlen.net}
\altaffiltext{1}{Institut d'Astrophysique de Paris - 98 bis boulevard Arago - 75014 Paris, France}
\altaffiltext{2}{The Johns Hopkins University, Department of Physics \& Astronomy, 3400 N. Charles St., Baltimore, MD 21218, USA}

\shorttitle{Cluster--Void Degeneracy Breaking}
\shortauthors{Sahl\'en, Zubeld\'ia \& Silk}

\begin{abstract}
Combining galaxy cluster and void abundances breaks the degeneracy between mean matter density $\Omega_{\rm m}$ and power spectrum normalization $\sigma_8$. In a first for voids, we constrain $\Omega_{\rm m} = 0.21 \pm 0.10$ and $\sigma_8 = 0.95 \pm 0.21$ for a flat $\Lambda$CDM universe, using extreme-value statistics on the claimed largest cluster and void. The {\it Planck}-consistent results detect dark energy with two objects, independently of other dark energy probes.  Cluster--void studies also offer complementarity in scale, density, and non-linearity -- of particular interest for testing modified-gravity models. 
\end{abstract}

\keywords{cosmological parameters --- cosmology: theory --- dark energy --- galaxies: clusters: individual (ACT-CL J0102-4915) --- large-scale structure of universe --- methods: statistical}

\maketitle

\section{INTRODUCTION}
\begin{figure}[t]
\includegraphics[width=\textwidth]{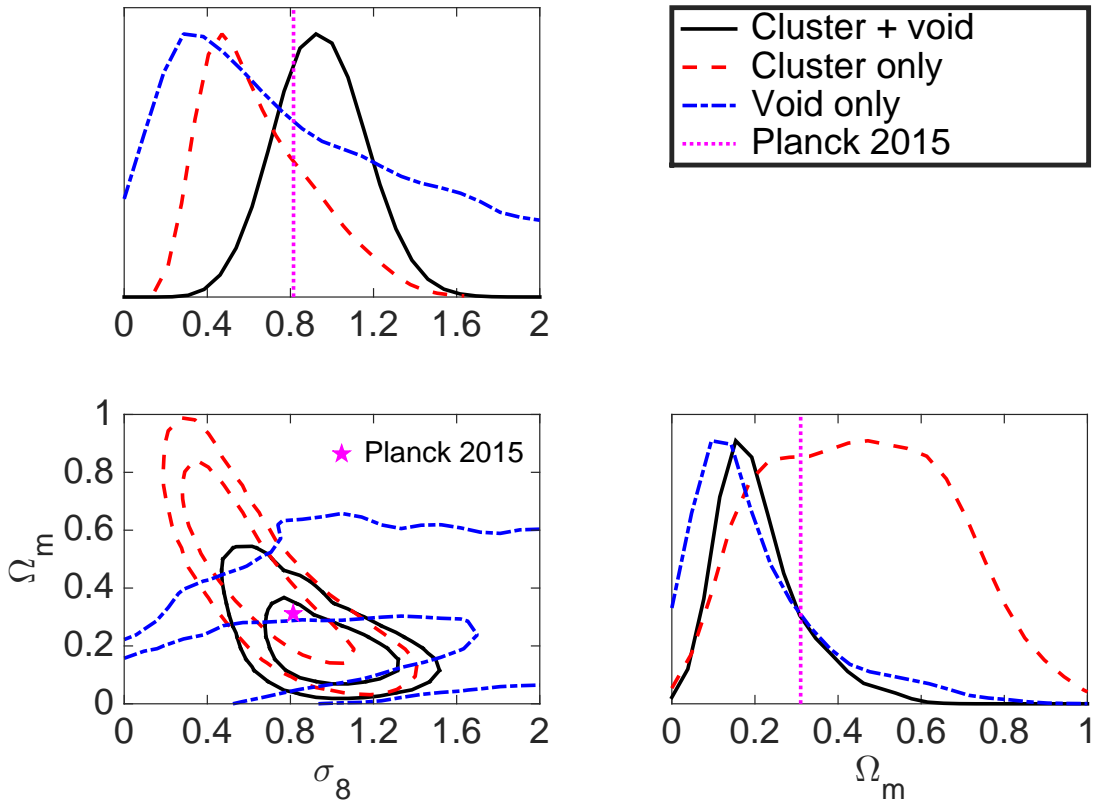}
\caption{ \label{omms8} Survey-patch constraints on $\Omega_{\rm m}$ and $\sigma_8$ ($68\%$ and $95\%$ confidence levels) from the largest cluster and void individually (red dashed, blue dash-dotted lines) and jointly (black solid lines).}
\end{figure}

Clusters and voids in the galaxy distribution are rare extremes of the cosmic web. As sensitive probes of the statistics of the matter distribution, they are useful tools for testing cosmological models. 
The abundances of clusters and voids are sensitive probes of dark energy \citep{2015arXiv150307690P, 2011ARA&A..49..409A}, modified gravity \citep{2015MNRAS.450.3319L,2011ARA&A..49..409A}, neutrino properties \citep{2010JCAP...09..014B,2015arXiv150603088M}, and non-Gaussianity \citep{2010ApJ...724..285C}.

Tests of the consistency of the most massive clusters with concordance $\Lambda$CDM cosmology have been performed over the last 20 years \citep{1998ApJ...504....1B,
2011PhRvD..83j3502H,2011PhRvD..83b3015M,
Harrison2013}.
No statistically significant tension prevails between the existence of the most massive clusters and concordance cosmology.  Large and deep voids have been considered as a possible problem for concordance cosmology \citep{
2001ApJ...557..495P,
2014MNRAS.441..933X,2015JCAP...05..062C}. While there are indications of possible discrepancies with $\Lambda$CDM, the significance of these is unclear. Only recently, sufficiently large and deep galaxy surveys enabled study of the statistics of the largest voids \cite[e.g.][]{2012ApJ...761...44S,2014MNRAS.440.1248N}.

\citet{2015JCAP...05..062C} studied the expected largest voids in concordance cosmology, and concludes some tension between expectation and observation. The use of void abundances to constrain $\sigma_8$ and $\Omega_{\rm m} h$ was discussed by \citet{2009MNRAS.400.1835B}, but they did not derive constraints, nor study the complementarity with cluster abundances. \citet{2015arXiv150307690P} explored future dark energy constraints from void abundances, and qualitatively argued for complementarity with cluster abundances. A joint cluster-void analysis and real-data cosmological parameter inference based on void abundance is still outstanding.

This work evaluates the complementarity of cluster and void abundances as cosmological probes, through analyzing whether the existence of the largest cluster and void is individually and/or jointly consistent with {\it Planck} cosmology \citep{Planck2015}. This includes the first quantitative cosmological parameter estimation based on void abundances. We find that cluster and void abundances powerfully break each other's parameter degeneracies, similar to complementarity between clusters and the cosmic microwave background (CMB). We explicitly model the effect of massive neutrinos, and Eddington bias in observables.

\section{DATA}
\label{sec:data}
We use data for the claimed largest galaxy cluster and void found so far. 
The surveys are summarized in Table~\ref{tab:surveys}. We also investigate an `observable Universe' case covering $z=0-6$.  

\begin{table}[htp]
\caption{Survey specifications.}
\label{tab:surveys}
\begin{center}
\begin{tabular}{|c|c|c|}
\hline
Survey & Area [sq. deg.] & Redshift \\
\hline
Cluster (ACT) & $1000$ & $0.3-6$ \\
Void ({\it WISE}--2MASS) & $21200$ & $0-0.3$  \\
\hline
\end{tabular}
\end{center}
\end{table}%

\subsection{Cluster}
A handful of galaxy clusters are `most massive known' candidates. We choose ACT-CL J0102-4915 `El Gordo' \citep{Menanteau2012}, which has a weak-lensing mass estimate that spans the range of uncertainty for the candidate set. It is most massive for at least $z > 0.6$, and its equivalent redshift-zero mass is most extreme \citep{Harrison2013}. 

`El Gordo' was discovered by the Atacama Cosmology Telescope (ACT) collaboration at a spectroscopic redshift $z=0.87$ through its Sunyaev--Zel'dovich (SZ) signal. The survey area is 1000 sq. deg. \citep{2013JCAP...07..008H}. It is estimated complete for $M_{200}~>~8~\times~10^{14} h^{-1}M_{\odot}$ in $z = 0.3-6$ \citep{Harrison2013}. The strongest SZ decrement in the survey patch corresponds to the cluster \citep{2013JCAP...07..008H}. We assume it is the most massive halo in the survey patch. We use the \textit{Hubble Space Telescope} weak-lensing mass measurement by \citet{Jee2014}, $M_{200} = \left(2.19 \pm 0.78\right) \times 10^{15} h^{-1}M_{\odot}$ (incl. syst. unc.).

\subsection{Void}
A supervoid (quasi-linear void) claimed the largest known was recently identified at $z = 0.22 \pm 0.03$, as a galaxy underdensity in the {\it WISE}--2MASS infrared galaxy catalogue \citep{Szapudi2014}. The void is aligned with the Cold Spot (CS) in the CMB \citep{Finelli2014}, hence we call it the `CS Void'. The survey area is 21200 sq. deg. \citep{Kovacs2014}, across $z=0-0.3$ \citep{2014arXiv1406.3622S}. No underdensity of larger size was identified in the survey \citep{2014arXiv1406.3622S}. We assume it is the largest void in the survey patch. \citet{Szapudi2014} measure a comoving radius $R = \left(220 \pm 50\right)h^{-1}$Mpc (in galaxy density) and a cold-dark-matter density contrast $\delta^{\rm cdm}_{\rm v} = -0.14 \pm 0.04$, after accounting for galaxy bias. The assumed fiducial $\Lambda$CDM model is $\Omega_{\rm m} = 0.3, \Omega_{\Lambda} = 0.7, h = 0.7$ \citep{KovacsPersComm}. We rescale the observed radius as $R = R_{\rm fiducial} [(dV/dz) / (dV/dz)_{\rm fiducial}]^{1/3}$ \citep{2015arXiv150307690P}. Void parameters are defined for a real-space top-hat model. We assume the survey is complete for voids of similar or larger size \citep{Kovacs2014}. For a large void partly contained in the survey, the density profile will be misestimated, but should be detected. A void covering the survey volume may be undetectable, but such voids ($R > 600 \,h^{-1}$~Mpc) should occur once in $>10^8$ Hubble volumes. Void numbers are exponentially suppressed with radius. Hence, Eq.~(\ref{eq:numbercount}), below, is insensitive to selection for voids much larger than the lower integration limit. Voids less underdense than the CS Void are not relevant for our analysis.

\section{METHOD}
\subsection{Model}
We predict cluster and void abundances adapting standard methodology \citep{Sahlen2009}. 

\subsubsection{Cosmological Model}
We assume a flat $\Lambda$CDM model with a power-law power spectrum of primordial density perturbation, and massive neutrinos. The model is specified by today's values of mean matter density $\Omega_{\rm m}$, mean baryonic matter density $\Omega_{\rm b}$, sum of neutrino masses $\Sigma m_{\nu}$, statistical spread of the matter field at quasi-linear scales $\sigma_8$, and scalar spectral index $n_{\rm s}$. We assume $\Sigma m_{\nu} = 0.06\,{\rm eV}$, one neutrino mass eigenstate and three neutrino species so that the effective relativistic degrees of freedom $N_{\rm eff} = 3.046$.

\subsubsection{Number Counts}
The model for number counts is
\begin{equation}
\label{eq:numbercount}
N_{\rm obs} = \int \int \int p(O | O_{\rm t}) n[M(O_{\rm t}), z]  \frac{dM}{dO_{\rm t}} \frac{dV}{dz} dz dO_{\rm t} dO,
\end{equation}
where $O$ is the size observable (mass, radius) for a type of object (here cluster or void), $O_{\rm t}$ the true physical value of the observable $O$, and $M(O_{\rm t})$ the mass of the object. The differential number density is given by $n(M,z)$, $p(O | O_{\rm t})$ is the measurement pdf for the observable $O$, and $dV/dz$ is the cosmic volume element. For integrating Eq.~(\ref{eq:numbercount}), we use $M_{\rm void}=\frac{4}{3}\pi R^3\rho_{\rm m}(1+\delta^{\rm m}_{\rm v})$. We highlight that while we write the expression for voids in terms of a mass, they are observationally defined by radius and density contrast. Redshift integration is performed according to survey specifications (Table~\ref{tab:surveys}), or `observable Universe' specification.  

\subsubsection{Number Density}
The differential number density of objects in a mass interval ${\rm d}M$ about $M$ at redshift $z$ is
\begin{equation}
\label{eq:numdens}
n(M, z)\,{\rm d}M = -F(\sigma, z)\,\frac{\rho_{\rm m}(z)}{M\sigma(M, z)}\,
\frac{{\rm d}\sigma(M, z)}{{\rm d}M}\,{\rm d}M\,,
\end{equation}
where $\sigma(M, z)$ is the dispersion of the density field at some comoving scale $R_L=(3M/4\pi\rho_{\rm m})^{1/3}$, and
$\rho_{\rm m}(z) = \rho_{\rm m}(z=0)(1+z)^3$ the matter density. The expression can be written in terms of linear-theory radius $R_L$ for voids. The multiplicity function (MF) denoted $F(\sigma, z)$ is described in the following for clusters and voids. 

\subsubsection{Cluster MF} 
The cluster (halo) MF $F_h(\sigma)$ encodes the halo collapse statistics. We use the MF of \citet{Watson2013}, their Eqs.~(12)-(15), accurate to within $10\%$ for the regime we consider. Since our measured mass is defined at overdensity  $200$ (rather than 178), we convert to a $\Delta=200$ overdensity using a scaling relation, their Eqs.~(17)-(19) \citep{Watson2013}.

Neutrinos are treated according to \citet{2010JCAP...09..014B}. Neutrinos free-stream on cluster scales, and do not participate in gravitational collapse, so cluster masses should be rescaled: $M = 4\pi R^3_L[(1-f_{\nu})\rho_{\rm m}+f_{\nu}\rho_{\rm b}]/3$, where $f_{\nu} = [\Sigma m_{\nu} / 93\,{\rm eV}] / \Omega_{\rm dm}h^2$ is the fraction of dark matter abundance $\Omega_{\rm dm}$ in neutrinos, and $\rho_{\rm b}$ is the baryon density. This gives a good first-order approximation, used with massless-neutrino MFs. Massive neutrinos also affect cluster and void distributions by shifting the turn-over scale in the matter power spectrum, and suppression of power on the neutrino free-streaming scale.

\subsubsection{Void MF}
The theoretical description of the void MF is not robustly known, mainly due to ambiguities in void definition, dynamics, and selection \cite[e.g.][]{2015JCAP...05..062C}. We use the Sheth--van de Weygaert (SvdW) MF \citep{Sheth2004} based on excursion set theory, but adapt it to supervoids, to describe a first-crossing distribution for voids with a particular density contrast $\delta_{\rm v}$. Void-in-cloud statistics are unimportant for our size of void. We pick the SvdW MF as representative also of alternative void MFs for our range of parameter values. 

\citet{2015MNRAS.451.3964A} show that several spherical-expansion + excursion set theory predictions of the void MF consistently describe the dark-matter void MF from $N$-body simulations within simulation uncertainties across at least $R=1-10\,h^{-1}$ Mpc. The SvdW MF is defined for a top-hat filter in momentum space. The observation is defined in real space, which introduces non-Markovian corrections. \citet{2015MNRAS.451.3964A} derive a `DDB' void MF including such corrections, and stochasticity in the void formation barrier. We have computed that these corrections give a $\sim 10\%$ increase in the void MF compared to SvdW for $R>100\,h^{-1}$~Mpc and $\delta_{\rm v} > -0.3$. The SvdW MF predicts $\mathcal{O}(1)$ voids with $R \sim 200\,h^{-1}$~Mpc in the {\it WISE}--2MASS survey patch, consistent with the peaks-theory prediction by \citet{2014PhRvD..90j3510N} rescaled to the survey patch, and extreme-void statistics in the largest $N$-body simulations \cite[e.g.][]{2015MNRAS.446.1321H}. Large, small-underdensity voids should be close to linear and trace  initial underdensity peaks, and hence not so sensitive to expansion dynamics and interactions. Indeed, the excursion-set-peaks prediction \citep{2012MNRAS.426.2789P} falls inbetween the SvdW and DDB MFs. Thus, SvdW should be a good approximation to the dark-matter void MF to within $\lesssim 50\%$. This uncertainty is significantly smaller than the Poisson uncertainty and propagated uncertainty in radius. These considerations motivate interest in a supervoid, rather than fully non-linear void.

While there is no consensus on the normalization of the dark-matter void MF, its shape is relatively well-understood for large supervoids. The same applies to voids defined by biased tracers such as galaxies, provided $\delta_{\rm v}$ is calibrated to survey specifications \citep{2015arXiv150307690P}. The radius and density contrast defined from tracers are larger than corresponding dark-matter quantities. Since we use a dark-matter density contrast, but a galaxy-density radius, we choose a converse approach. We set $\delta_{\rm v}$ to the measured dark-matter value (within uncertainties), and allow the radius to be self-calibrated \citep{2003PhRvD..67h1304H} to a dark-matter radius by the data. This procedure is successful in that only void radius but not density contrast is re-calibrated in parameter estimation. This is analogous to the scaling-relation strategy used for galaxy clusters \citep{2011ARA&A..49..409A}. 

The void MF for (non-linear) radius $R$ should be evaluated at corresponding linear radius $R_{\rm L}$, related as $R/R_{\rm L} = (1+\delta^{\rm m}_{\rm v})^{-1/3}$, with $\delta^{\rm m}_{\rm v}$ non-linear. The linear density contrast $\delta_{\rm L}$ defines the MF density threshold. We approximate the spherical-expansion relationship by $\delta^{\rm m}_{\rm v, L} = c[1-(1+\delta^{\rm m}_{\rm v})^{-1/c}]$, $c=1.594$ \citep{Jennings2013}. 

Neutrinos are treated according to \citet{2015arXiv150603088M}. Neutrino density contributes to the dynamical evolution of the void, but lacks significant density contrast itself. We use $\delta^{\rm m}_{\rm v} = [1-f_{\nu}(1+\Omega_{\rm b}/\Omega_{\rm m})]\delta^{\rm cdm}_{\rm v}$, where $\delta^{\rm cdm}_{\rm v}$ is the {\it cold-dark-matter} density contrast (which is our observational quantity). 

\subsubsection{Measurement Uncertainty}
We use a log-normal pdf $p(O | O_{\rm t})$ in Eq.~(\ref{eq:numbercount}) for the observable $O$ (i.e. $M_{200}$ or $R$) given its true value $O_{\rm t}$, with the mean and variance matched to the mean and variance of the observational cluster mass or void radius determinations, $\mu_{\ln O} = \ln O_{\rm t}/\sqrt{1+(\sigma_O/O_{\rm t})^2}$ and $\sigma^2_{\ln O} = \ln[1+(\sigma_O/O_{\rm t})^2]$. This approximation does not bias predicted abundances. Including this probability distribution in predictions corrects for the otherwise-present Eddington bias. We also perform an Eddington-biased analysis for which $p(O | O_{\rm t}) = \delta(O - O_{\rm t})$. 
Redshift uncertainties are neglected, as they are contained in the survey patches. 

\subsection{LIKELIHOOD}
\subsubsection{Cluster--Void Extreme Value Statistics}

Number counts follow Poisson statistics, corrected for object-to-object clustering. The fractional correction to Poisson counts due to mild $[N_{\rm c}(>M) \ll 1]$ clustering is $-N_{\rm c}(>M)/2$ \citep{Colombi2011}, where $N_{\rm c}(>M) \equiv  \bar{\xi}(>M) N(>M)$, and $\bar{\xi}(>M)$ is the average auto- or cross-correlation for objects above the mass threshold(s) within the patch. For all our cluster and void cases, we estimate $N_{\rm c}(>M) \lesssim 0.01$, based on \cite{Davis2011}, and correlation function and bias results in \cite{Sheth2004,2014PhRvL.112d1304H,2016MNRAS.456.4425C}. Hence, our number counts follow non-clustered Poisson statistics to sub-percent-level accuracy.

Extreme value (or Gumbel) statistics describe the pdf of extrema of samples drawn from random distributions \citep{Gumbel1958}. Consider a large patch of the Universe, populated with clusters and voids. Let $M_{\rm max}$ be the mass of the most extreme cluster (or void), and  $p_{\rm G}\left(M_{\rm max}\right)$ be the pdf of $M_{\rm max}$. For non-clustered Poisson statistics, we can write \citep{Davis2011,Colombi2011}:
\begin{equation} \label{eq:likelihood}
p_{\rm G}\left(M\right) = \frac{dN(>M)}{dM}e^{-N(>M)}\,,
\end{equation}
where $N(>M)$ is the mean number of clusters (or voids) above the threshold $M$ within the patch, according to Eq.~(\ref{eq:numbercount}). 
The total Gumbel likelihood (under non-clustering) is then
\begin{equation}\label{eq:like}
p_{\rm G}(M_{\rm halo}, M_{\rm void}) = p^{\rm halo}_{\rm G}\left(M_{\rm halo} \right)p^{\rm void}_{\rm G}\left(M_{\rm void} \right)\,.
\end{equation}
This likelihood is multiplied with measurement pdfs and external priors, and evaluated as described in Sec.~\ref{sec:mcmc}.

\subsubsection{Measurement Uncertainty}
We include measurement uncertainties in cluster and void properties as normally distributed with means and variances as detailed in Sec.~\ref{sec:data}. 

\subsubsection{External Priors} 
In line with recent {\it Planck} analyses \citep{Planck2015}, we use the following external priors: 
\begin{align}
h & = 0.706 \pm 0.033\,\,\text{\rm (maser+Cepheids,\,\citealt{2014MNRAS.440.1138E})}\,, \\ 
\Omega_{\rm b} h^2 & = 0.023 \pm 0.002\,\,\text{\rm (BBN\footnotemark,\,\citealt{Agashe:2014kda})} \,, \\ 
n_{\rm s} & = 0.9677 \pm 0.006\,\,\text{\rm (CMB\footnotemark,\,\citealt{Planck2015})} \,, \\ 
\Sigma m_{\nu} & = 0.06\,\text{\rm eV\,(1\,mass\,eigenstate)\,\,(NO\footnotemark,\, \citealt{Agashe:2014kda})} \,, \\ 
 N_{\rm eff} & = 3.046\,\,\text{\rm (Std.\,Model, 3\,neutrino\,species)}\,.
\end{align}
\addtocounter{footnote}{-2}
\footnotetext{Big Bang nucleosynthesis}
\stepcounter{footnote}
\footnotetext{Cosmic microwave background anisotropies}
\stepcounter{footnote}
\footnotetext{Neutrino oscillations}
We specify $T_{\rm CMB} = 2.7255$ K \citep{2009ApJ...707..916F} for the radiation energy density. With these priors, our analysis is directly comparable to the {\it Planck} results. For an equivalent analysis independent of CMB anisotropies, we can set $n_{\rm s} \approx 1$ assuming an inflationary origin of primordial perturbations, or based on large-scale-structure surveys.

\subsection{Computation}
\label{sec:mcmc}
We perform Monte Carlo Markov Chain parameter estimation on the parameter space $\{h, \Omega_{\rm m}, \Omega_{\rm b}, \sigma_8,$ $n_{\rm s}, R_{\rm eff}, \delta^{\rm cdm}_{\rm v}, M_{200}\}$, where $R_{\rm eff}$ is the effective self-calibrated dark-matter void radius. The likelihood, Eq.~(\ref{eq:like}) $\times$ measurement pdfs $\times$ priors, is evaluated using a modified version of CosmoMC \citep{Lewis2002}, employing techniques in \citet{Sahlen2009}. Abundances are computed using CUBPACK \citep{Cubpack}.

\section{RESULTS}
\subsection{Number Counts}
In a {\it Planck} cosmology, we predict in the survey patches an Eddington-biased (unbiased) mean number of clusters $N(>M_{200})$ in the range $[10^{-2}, 87]\,([10^{-4}, 49])$, and mean number of voids $N(>R)$ in the range $[10^{-11}, 22]\,([10^{-100}, 20])$. These correspond to $2\sigma$ CIs of cluster and void properties.

Massive neutrinos produce a relatively insignificant suppression in mean {\it Eddington-biased} numbers. The unbiased means in the above cases are suppressed between $\sim 5-50\%$. 

\subsection{Cosmological Parameters}
Parameter contours for $\Omega_{\rm m}$ and $\sigma_8$ are shown in Fig.~\ref{omms8}, and marginalized constraints in Table~\ref{tab:params}. {\it Planck} concordance cosmology is inside the $68\%$ confidence contours for the cluster-only and cluster+void cases, and the $95\%$ contours for the void-only case. The objects are individually and mutually consistent with concordance cosmology. The joint analysis detects dark energy at $>7\sigma$.  

\begin{table}[htp]
\caption{Marginalized means for $\Omega_{\rm m}$ and $\sigma_8$ (with $68\%$ CIs), and best-fit values, in the joint cluster--void analysis.}~
\label{tab:params}
\begin{center}
\begin{tabular}{|c|c|c|c|c|}
\hline
& \multicolumn{2}{|c|}{Eddington-corrected} & \multicolumn{2}{|c|}{Eddington-biased} \\
\hline
Case & $\Omega_{\rm m}$ & $\sigma_8$ & $\Omega_{\rm m}$ & $\sigma_8$ \\
\hline
Survey & 
  \begin{tabular}{@{} c @{}}
    $0.21 \pm 0.10$ \\ 
    $0.27$ \\ 
  \end{tabular}
 & 
  \begin{tabular}{@{} c @{}}
    $0.95 \pm 0.21$ \\ 
    $0.82$ \\ 
  \end{tabular} 
  & 
    \begin{tabular}{@{} c @{}}
    $0.22 \pm 0.12$ \\ 
    $0.15$ \\ 
  \end{tabular} 
   & 
    \begin{tabular}{@{} c @{}}
   $0.99 \pm 0.22$\\ 
    $1.17$ \\ 
  \end{tabular}    
    \\
 \hline
     \begin{tabular}{@{} c @{}}
   Observable Universe \\ 
   $z=0 - 6$\\ 
  \end{tabular}   
 & 
    \begin{tabular}{@{} c @{}}
   $0.23 \pm 0.11$\\ 
    $0.32$ \\ 
  \end{tabular}    
 & 
    \begin{tabular}{@{} c @{}}
    $0.70 \pm 0.14$ \\ 
    $0.72$ \\ 
  \end{tabular}    
 & 
    \begin{tabular}{@{} c @{}}
 $0.21 \pm 0.08$\\ 
    $0.20$ \\ 
  \end{tabular}    
 &
    \begin{tabular}{@{} c @{}}
  $0.78 \pm 0.13$\\ 
    $0.80$ \\ 
  \end{tabular}     
  \\
\hline
\end{tabular}
\end{center}
\end{table}%

\subsection{Self-calibration} 
The void radius is consistently self-calibrated to a dark-matter-field radius of $R_{\rm eff} = 184 \pm 51\,h^{-1}$ Mpc in the Eddington-corrected survey-patch analysis, with the density-contrast pdf preserved. 

\subsection{Eddington Bias}
Eddington bias does not significantly affect the constraints. The bias, if unaccounted for, is a slight shift in $\Omega_{\rm m}$ and of around $0.5\sigma$ in $\sigma_8$, see Table~\ref{tab:params}.

\subsection{Degeneracy}
The abundance of clusters and voids generate mutually orthogonal degeneracies in the $\Omega_{\rm m}-\sigma_8$ plane. This is due to the different ways in which scales $R_{\rm L}$ are determined: for clusters, $R_{\rm L}$ is derived from the measured mass and $\Omega_{\rm m}$; for voids, $R_{\rm L}$ is derived from the measured angular/radial extent and depends geometrically on $\Omega_{\rm m}$ \citep{2015arXiv150307690P}. Degeneracies are also determined by sensitivity to proper distance (volume increase) and power-spectrum normalization and growth. Proper distance dominates at low redshift, growth at high redshift \citep{2002ApJ...577..569L}. For flat $\Lambda$CDM, proper distance and growth are both fixed by $\Omega_{\rm m}$. Constant proper distance and constant growth coincide in parameter space for all redshifts, and cluster--void degeneracy should change little with redshift. We checked that degeneracy for average-size voids coincide for $z = 0 - 0.3$ and $z = 0.4 - 0.7$. Orthogonality remains for shell-crossed voids ($\delta^{\rm m}_{\rm v} \approx -0.8$). More general models will have redshift-dependent degeneracies. Clusters are growth-sensitive for $z \gtrsim 0.5$ \citep{2002ApJ...577..569L}. Voids form slower, and are therefore sensitive to perturbation growth at low redshifts: for shell-crossed average-size voids at $z\gtrsim0.2$, for $\delta^{\rm m}_{\rm v} = -0.5$ at $z\gtrsim 0.5$. Voids like the CS Void are growth-sensitive for $z\gtrsim 0.1$. Complementarity with clusters is thus robust. {\it Euclid} clusters and voids are complementary in the dark energy equation-of-state $w_0$--$w_a$ parameter plane (cf. Fig.~4 in \citealt{2015arXiv150502165S}, Figs.~4 \& 6 in \citealt{2015arXiv150307690P}).

\section{CONCLUSIONS}
The claimed largest galaxy cluster and void are individually consistent with {\it Planck} concordance $\Lambda$CDM cosmology, regardless of whether constrained to the respective survey patches or the full observable Universe (and regardless of Eddington bias). This relaxes the $\sim 3\sigma$ significance of the primordial fluctuation \citep{Szapudi2014}, and is consistent with predictions by \citet{2014PhRvD..90j3510N}. \citet{2015JCAP...05..062C} notes, consistent with our findings, that large voids typically need small density contrast to be consistent with concordance cosmology. The cluster and void are jointly consistent with {\it Planck}, preferring a somewhat low value of $\Omega_{\rm m} = 0.21 \pm 0.10$. This is a `pure' large-scale-structure detection of dark energy at $>7\sigma$ based on only two data points (covering linear $k \sim 0.01 - 1\,h$ Mpc$^{-1}$, $\delta \sim -0.2 - 2$, factor $10^8$ in density and gravitational potential) plus local-Universe and particle-physics priors. The detection is independent of other dark-energy probes, e.g. CMB anisotropies, type Ia supernovae, baryon acoustic oscillations, and redshift-space distortions.  Better data on cluster mass and void radius could reveal a tension with {\it Planck}. A higher-redshift (e.g. $z>0.5$) void survey could provide stronger parameter constraints, particularly for more general cosmological models. This is due to greater sensitivity to perturbation growth at higher redshifts, and greater statistical weight of large, early-formed objects. Our model shows that single large high-redshift voids can have the same statistical weight as a few handfuls of low-redshift voids of similar size.

The main possible systematics are associated with cluster and void MFs, and measurements of void properties. The theoretical MF uncertainty of around $10-50\%$ is subdominant to the Poisson uncertainty of $\mathcal{O}(1)$ objects. This is corroborated by finding no significant difference in constraints regardless of patch specifications or Eddington bias. Measurements of void properties may suffer from systematics associated with ellipticity. Voids are typically elliptical, thus the top-hat radius could be biased. A typical ellipticity of $15\%$ \citep{2015JCAP...03..047L} corresponds to a $\leq 40\%$ line-of-sight deviation in the extent of the void relative to the measured radius. Such a bias is smaller than the quoted CS Void radius uncertainty. The CS Void integrated Sachs--Wolfe (ISW) decrement can not as-is explain the Cold Spot  \citep{2014PhRvD..90j3510N}. The void radial profile is also poorly constrained on the far side \citep{Szapudi2014}. These observations might suggest that the CS Void is elliptical along the line of sight. This is a subject for follow-up measurements. However, even with significant ellipticity, the void would not explain the Cold Spot ISW decrement \citep{2015arXiv151009076M}. 

Our conclusions are robust with respect to the possibility that `El Gordo' might not be the most massive cluster in `survey' or `observable Universe'. Its mass estimate accounts for the mass uncertainty within both `most massive' candidate sets. Only `survey' or `observable Universe' redshift ranges enter the analysis.

The CS Void density contrast measurement allows a $0.05\%$ probability of no underdensity ($\delta_{\rm v} \geq 0$). This, or other CS Void question marks, can be relieved for our analysis by instead considering the Local Void. Measurements of the local galaxy luminosity density indicate that the local Universe coincides with an extreme void. Based on \citet{2013ApJ...775...62K, 2014arXiv1409.8458K}, we estimate its top-hat galaxy-density radius $R \approx 210\,h^{-1}$~Mpc and dark-matter density contrast $\delta_{\rm v}^{\rm cdm} \approx -0.15 - -0.2$. The values are very close to those of the CS Void. Since both voids coincidentally prescribe very similar largest-void properties, our analysis is approximately independent of choice of void. The Local Void is consistent with concordance cosmology \citep{2014MNRAS.441..933X}. 

We identify a powerful complementarity between cluster and void abundance constraints on the mean matter density and  matter power spectrum. The void-abundance degeneracy and comoving-scale sensitivity is similar to that of CMB anisotropies. Joint cluster--void abundances may therefore provide strong constraints on, e.g., the matter power spectrum, neutrino properties, dark energy, and modified gravity. The cluster--void complementarity in scale and density is compelling for screened theories of gravity.  We will investigate these possibilities in a follow-up publication.

Surveys with {\it XMM--Newton}\footnote{www.xcs-home.org, irfu.cea.fr/xxl}, {\it eROSITA}\footnote{www.mpe.mpg.de/eROSITA}, the Dark Energy Survey\footnote{www.darkenergysurvey.org}, the Dark Energy Spectroscopic Instrument\footnote{desi.lbl.gov}, {\it Euclid}\footnote{www.euclid-ec.org}, the Square Kilometre Array\footnote{www.skatelescope.org}, the 4-metre Multi-Object Spectroscopic Telescope\footnote{www.4most.eu}, the Large Synoptic Survey Telescope\footnote{www.lsst.org}, and the {\it Wide-Field Infrared Survey Telescope}\footnote{wfirst.gsfc.nasa.gov}, will yield tens to hundreds of thousands of clusters and voids -- for a factor $10^2 - 10^3$ tighter constraints. Joint cluster--void analyses promise to be a powerful cosmological probe and tool for systematics calibration, where the novel complementarity in degeneracy, scale and density regimes can constrain cosmology beyond the concordance model. 

\acknowledgments
We thank J.~Dunkley, A.~Goobar, J.~Hughes, M.~J.~Jee, E.~Jennings, A.~Kov\'acs, E.~M\"ortsell, I.~Szapudi and H.~Winther for helpful discussions, and the anonymous referee for well-received suggestions for clarifications. MS and IZ were supported by the Templeton Foundation. JS acknowledges support by ERC project 267117 (DARK) hosted by UPMC, by NSF grant OIA-1124403 at JHU, and by the Templeton Foundation. This work was undertaken on the COSMOS Shared Memory system at DAMTP, University of Cambridge operated on behalf of the STFC DiRAC HPC Facility. This equipment is funded by BIS National E-infrastructure capital grant ST/J005673/1 and STFC grants ST/H008586/1, ST/K00333X/1.

{\it Facilities:} \facility{ACT}, \facility{CTIO:2MASS}, \facility{HST}, \facility{WISE}.

\end{document}